\documentclass[aps,prd,floatfix,onecolumn,superscriptaddress,preprintnumbers,nofootinbib,10pt]{revtex4}
\usepackage{slashed}
\usepackage{epsfig,latexsym,cancel,amssymb,amsmath,verbatim,mathrsfs}
\usepackage{ulem}
\usepackage{color}
\definecolor{My_red}        {cmyk}{0.00,1.00,1.00,0.20}

\usepackage{graphicx}

\def\ra{\rightarrow}
\def\L{\left(}
\def\R{\right)}
\def\wt{\widetilde}
\def\Ld{\Lambda}
\def\ld{\lambda}
\def\f{\frac}
\newcommand{\be}{\begin{equation}}
\newcommand{\ee}{\end{equation}}
\newcommand{\bea}{\begin{eqnarray}}
\newcommand{\eea}{\end{eqnarray}}
\newcommand{\ba}{\begin{array}}
\newcommand{\ea}{\end{array}}

\long\def\symbolfootnote[#1]#2{\begingroup%
\def\thefootnote{\fnsymbol{footnote}}\footnote[#1]{#2}\endgroup}

\newcommand{\beq}{\begin{equation}}
\newcommand{\eeq}{\end{equation}}

\begin{document}

\title{Higgs Naturalness and Dark Matter Stability by Scale Invariance}


\author{Jun Guo}
\email[E-mail: ]{hustgj@itp.ac.cn}
 \affiliation{State Key Laboratory of
Theoretical Physics and Kavli Institute for Theoretical Physics
China (KITPC), Institute of Theoretical Physics, Chinese Academy of
Sciences, Beijing 100190, P. R. China}


\author{Zhaofeng Kang}
\email[E-mail: ]{zhaofengkang@gmail.com}
\affiliation{School of Physics, Korea Institute for Advanced Study,
Seoul 130-722, Korea}
\affiliation{Center for
High-Energy Physics, Peking University, Beijing, 100871, P. R.
China}

\date{\today}

\begin{abstract}

Extending the spacetime symmetries of standard model (SM) by scale invariance (SI) may address the Higgs naturalness problem. In this article we attempt to embed accidental dark matter (DM) into SISM, requiring that the symmetry protecting DM stability is accidental due to the model structure rather than imposed by hand. In this framework, if the light SM-like Higgs boson is the pseudo Goldstone boson of SI spontaneously breaking, we can even pine down the model, two-Higgs-doublets plus a real singlet: The singlet is the DM candidate and the extra Higgs doublet triggers electroweak symmetry breaking via the Coleman-Weinberg mechanism; Moreover, it dominates DM dynamics. We study spontaneously breaking of SI using the Gillard-Weinberg approach and find that the second doublet should acquire vacuum expectation value near the weak scale. Moreover, its components should acquire masses around 380 GeV except for a light CP-odd Higgs boson. Based on these features, we explore viable ways to achieve the correct relic density of DM, facing stringent constraints from direct detections of DM. For instance, DM annihilates into $b\bar b$ near the SM-like Higgs boson pole, or into a pair of CP-odd Higgs boson with mass above that pole.

\end{abstract}

\maketitle

\section{Introduction and motivation}

Recently, the LHC discovered a new resonance around 125 GeV, which is putative the Higgs boson predicted by the standard model (SM)~\cite{Higgs:dis}. Thus far, its measured couplings are well consistent with the SM predictions and give no illustrative hints for new physics beyond the SM. But the existence of a light spin-0 particle at low energy has already given an important hint to new physics: a fundamental scalar boson suffers the notorious hierarchy problem which must be addressed by new physics. As a matter of fact, it has been guiding the direction of going beyond SM in decades. The well accepted solutions can be classified into two categories: One is canceling the quadratic divergency by virtue of symmetries such as supersymmetry and the other one is imposing a low cut-off scale around TeV, such as in the composite and large extra dimension models. But Bardeen in the paper Ref.~\cite{Bardeen} proposed a quite different solution, in which the classical scale invariance (SI) is supposed to protect the light Higgs field $\Phi$. SI may change our conventional understanding of the quadratic divergency, i.e., it is an artificial quantity in an improper regularization method (We will present the argument in the text). Although this viewpoint is in controversial, SI still deserves serious attention for its potential to be an economical solution to the big hierarchy problem~\cite{AlexanderNunneley:2010nw,Dermisek:2013pta}.

On the other hand, the existence of dark matter (DM) is also commonly believed to be a good guild for new physics (see some studies of DM within the SI framework~\cite{SI:DM}). Unfortunately, our knowledge of DM is not rich, and its particle properties such as spin, mass and interactions all are unknown. This leaves a huge room for DM model building, but the resulting predictions are in a mass. In this article we try to address one of the basic questions about DM, i.e., why is it stable?  As is well known, the baryon number that protects proton stability is an accidental symmetry, as a result of the gauge symmetries and field content of the SM. Inspired by this, we explore the idea of accidental DM (aDM) by virtue of the extended space-time symmetries of the scale invariant SM (SISM). It, then, amounts to asking what field content can be allowed. The answer is encouraging. Under some reasonable assumptions and aided by the current experimental data, we find that only a second Higgs doublet and a real singlet $S$ are allowed. Here the real singlet plays the role of aDM, while the second doublet triggers electroweak symmetry breaking (EWSB) via the Coleman-Weinberg (CW) mechanism~\cite{CW}. Furthermore, the second Higgs doublet furnishes key elements for the thermal aDM to acquire correct relic density. In our focused case, it presents a naturally light CP-odd Higgs boson $A$, into which aDM can annihilate without incurring a large DM-nucleon recoil rate (but with fine-tuning to make DM and $A$ almost degenerate). Remarkably, the model makes some testable predictions: The aDM should lie near or above half of the SM-like Higgs boson mass but lighter than 100 GeV, and it will be found or ruled out by the next round of LUX; The extra Higgs states have mass around 380 GeV except for the light $A$, and all of them can be hunted at LHC.

In the framework of aDM, we will ``derive" instead build the two Higgs doublet model (2HDM) plus a real singlet. This approach distinguishes our study from the previous relevant studies. Actually, the ordinary version of this model, i.e., that has no SI thus resorting to $Z_2$ protective symmetry on the singlet, has been studied in~\cite{Cai:2013zga}. The SI version of 2HDM with a second Higgs doublet triggering EWSB is considered in Ref.~\cite{1Dermisek:2013pta}. However, to make the extra doublet provide a DM candidate, authors again have to impose a $Z_2$ symmetry by hand. The SISM extended with a real singlet only is the simplest model that provides aDM. If it accommodates viable phenomenologies, i.e., the singlet could trigger EWSB and at the same time be a good DM, we should cheer for this model. Unfortunately, those two aspects are in so strong tension that it likely fails. In the completion of this article, a thorough study on this model, including higher order corrections, was made in Ref.~\cite{Steele:2013fka}. They found that even in the dynamical approach~\footnote{It is inspired by the method of E. Gildener and S. Weinberg~\cite{GW} to treat a scalar potential with multi Higgs fields. In this method the Higgs and singlet masses are generated simultaneously through radiative EWSB. But in this paper we will use the conventional sequential method, where the singlet gets mass after radiative EWSB.} , the singlet can only account for less than $26\%$ of the total DM relic density. 

The paper is organized as following. In Section II we discuss the relation between classical scale invariance and hierarchy problem. In addition, radiative SI spontaneously breaking is briefed. In Section III we embed aDM into SISM and establish 2HDM$+S$. Phenomenologies of the Higgs sector and dark matter are detailed. Discussion and conclusion are cast in Section IV. 
 
\section{Classical Scale invariance and Higgs naturalness}

In this Section we will first understand quadratic divergency and the related hierarchy problem in a nonconventional way, and then argue the possible role played by classical SI in solving this problem. General aspects of spontaneously breaking of SI, as a consequence of SI anomaly, are briefly reviewed.

\subsection{Classical Scale invariance and the hierarchy problem}

In quantum field theory, a scalar filed $\phi$ is expected to receive quadratic divergency in calculating radiative corrections on the mass term $m_\phi^2|\phi|^2$, with $m_\phi$ the renormalized mass parameter. This divergency is manifested in the cut-off regularization which isolates the divergent terms in the form of $y \Ld^2 |\phi|^2$, where $y$ denotes the dimensionless coupling between $\phi$ and the particle running in the loop. If $\Ld$ is much higher than the weak scale, then the hierarchy problem arises: How to naturally make the weak scale ($\sim m_\phi$) much below it? To handle this problem, it is important to identify the role of $\Ld$. If $\Ld$ is regarded as a physical mass scale, e.g., the mass of the loopy particle introduced to cancel the previous divergency or the composite scale above which the patrons of Higgs field take over the theory, then $\Ld$ should be around the TeV scale. Actually, the hierarchy problem was first raised in the context of Higgs coupling to heavy fields whose masses are identified with $\Ld$~\cite{hp}.

But what if there is not an explicit physical scale? This might be the worst case, because people take the assumption that any low energy theory will turn out to be invalid above the Plank scale and consequently $\Ld\sim M_{\rm Pl}$. It is supported by the phenomena in condense physics, e.g., in a magnetic system the atomic lattice spacing $a$ provides an ultraviolet (UV) cutoff scale on momentum $\Ld\sim a^{-1}$. However, we are not confident that the space-time we are living in is also a lattice. We neither do not have a confirmative way to include gravity, whose quantization is not known yet, in the standard model of particle physics, which is clearly described by quantum field theory. Therefore, it is not unreasonable to think that $\Ld$ is merely a technique tool in the cut-off like regularization methods. Eventually, the dependence on $\Ld$ will be removed in any renormalizable theory and never show in the physical quantities. In this sense, the quadratic divergency problem is a technique problem and can be technically solved through the renormalization procedure. The dimension regularization (DR) method~\cite{DR} strongly advocates this viewpoint, because in it quadratic divergency does not appear at all. The subtraction of quadratic divergency is also done in a massive $\phi^4$ model with a large fixed cut-off scale, using higher derivative regularization~\cite{Fujikawa} and Wilsonian renormalization~\cite{Aoki:2012xs}. In a word, although the renormalization of $m_\phi^2$ consists of the subtractive and multiplicative renormalizations, they can be separated; this fact justifies the unphysical face of quadratic divergency~\footnote{Authors of the latter paper further argued that the subtractive renormalization does not render fine-tuning, since the subtraction amounts to determining the position of the critical line (or,  quadratic divergency can be absorbed into the position of critical line).}.

The absence of an explicit UV physical scale is just the case in the scale invariant theories. By definition, a model with SI should contain no prior scale, even the Planck scale. Otherwise, it falls into the framework of Wilsonian effective field theory, in which the mass term is unavoidably generated after integrating out the higher momentum modes up to $M_{\rm Pl}$. Then, a light or massless scalar field can survive only as a result of complete quadratic divergency subtraction~\cite{Aoki:2012xs}. This means that the beginning theory at $M_{\rm Pl}$ must have a very large bare mass rather than massless required by SI. Now we arrive the conclusion, if the SM space-time symmetries are extended by SI, we in the loop calculation will have to either use DR or admit a tool-like $\Ld$. The latter is just the case in the original proposal of using classical SI to protect weak scale~\cite{Bardeen}, where the term $\Ld^2|\phi|^2$ is subtracted via renormalization condition. We would like to stress that, although the classical SI seems to be preserved in the cut-off like regularization after renormalization, in practice there SI is already violated at the stage of regularization. While DR does not introduce a scale in regularization and thus is free of violating SI via the quadratic divergency term. Of course, eventually DR introduces a mass scale during renormalization, which then leads to the anomaly of classical SI. But anomaly is not a disaster to the solving of hierarchy problem, because it only leads to logarithmically running of dimensionless couplings but never recurs quadratic divergency. As a matter of fact, this anomaly is a key to understand the origin of EW scale, which will be addressed soon later.

\subsection{Classical SI anomaly and the origin of EW scale} \label{CW:ge}

The SM has a characteristic scale, the EW scale, so in a realistic model SI should be broken somehow. Interestingly, as mentioned before the classical SI actually is broken by quantum effects (namely anomaly), which can cause SI spontaneously breaking (SISB) indeed. It is tempting to identify the SISB scale with the EW scale, since we then have an economical way to understand EWSB. Moreover, in this scenario the SM-like Higgs boson is a pseudo Goldstone boson (pGSB) of SISB, and hence its lightness is well understood. In fact,  symmetry spontaneously breaking in a scaleless theory was explored long ago by Coleman and Weinberg (CW) in their classical paper~\cite{CW}, and they found that it can happen through dimensional transmutation.

Let us briefly review how the CW mechanism works and its generic features. We confine to the one-dimension field space of a single classical field $\phi_{\rm cl}$, where the vacuum is determined by the minimum of the 1PI effective potential $V_{\rm eff}(\phi_{\rm cl})$. In the scaleless theory, at one-loop level $V_{\rm eff}(\phi_{\rm cl})$ can be generically written as
\begin{align}\label{veff}
V_{\rm eff}= {A} \phi_{\rm cl}^4+{B} \phi_{\rm cl}^4 \ln\f{\phi_{\rm
cl}^2}{Q^2}.
\end{align}
${A}$ and ${B}$ are functions of the dimensionless constants involving the couplings of $\phi_{\rm cl}$. In the $\overline {\rm MS}$ scheme, they are
\begin{align}\label{beta}
&{A}=\f{\ld_\phi}{8}+\f{1}{64\pi^2}\sum_P n_P g_{P}^4\L -A_P+\ln
{g_{P}^2} \R,\cr & {B}=\f{1}{64\pi^2}\sum_P n_P g_{P}^4,
\end{align}
with $\ld_\phi$ the tree-level quartic coupling constant (see Eq.~(\ref{singlet})). $P$ sums over particles which have internal degrees of freedom $n_P$ and field-dependent masses $m_P=g_P\phi_{\rm cl}$. The factor $A_P=3/2,\,3/2,\,5/6$ for the spin 0, 1/2 and 1 particles, respectively. Note that now an explicit scale $Q$, the renormalization scale introduced in DR, appears in the second term. It reflects anomaly of SI and is the key for SISB. An extremum at $\langle\phi_{\rm cl}\rangle$ is created given $\ln ({Q}/{\langle\phi_{\rm cl}\rangle})=\f{1}{4}+{{A}}/{2{B}}$. With it one can eliminate $Q$ in the potential, in favor $\langle\phi_{\rm cl}\rangle$, $A$ and $B$. Further expanding the potential around $\langle\phi_{\rm cl}\rangle$, it is not difficulty to get the curvature of $V_{\rm eff}$ at $\langle\phi_{\rm cl}\rangle$:
\begin{align}\label{SIPGSB}
m_\phi^2=8{B} \langle\phi_{\rm cl}\rangle^2.
\end{align}
If ${B}<0$ the extremum will be a maximum; If ${B}>0$ it will be a local minimum showing SISB. As mentioned previously, the resulting GSB is pseudo (because the classical SI is broken by anomaly), so its mass squared $m_\phi^2$ is loop suppressed but not massless. This fact helps us to understand the lightness of the SM-like Higgs boson when we identify it with this pGSB.

Applying the above generic analysis to the SM imposed with SI (without any field extension), in which the neutral CP-even component of the Higgs doublet $\Phi$ provides the classical field $\phi_{\rm cl}$, one soon finds that the model gives ${B}<0$ and therefore fails in triggering EWSB. The cause of the failure is evident: The observed top quark has a large mass and moreover its internal degrees of freedom is negatively big, $n_t=-12$. To overcome this problem, naturally we introduce scalar or vector bosons, which are capable of flipping the sign and making $B>0$, provided that they have fairly large couplings to $\Phi$~\cite{Dermisek:2013pta}. A variant is by means of the Higgs portal $\ld_X(\Phi^\dagger\Phi)|X|^2$, which can produce the ordinary negative Higgs mass term $m_{\Phi}^2=\ld_X v_X^2$ given $\ld_X<0$. Here $v_X=\langle X\rangle$ is by virtue of a hidden CW mechanism~\cite{AlexanderNunneley:2010nw} (or hidden confining gauge dynamics~\cite{Hur:2011sv}). The former approach accommodates a predictive framework for DM, so we concentrate on it in this paper. 

\section{Accidental dark matter (aDM) by SI}

DM guides us to new physics beyond the SM, and it inspires a pool of models with various motivations. However, most of them can not explain why DM is stable and thus an artificial symmetry that ensures DM stability should be imposed. This situation is absolutely different to that of the visible matters like proton, whose sufficient stability is ensured by the accidental baryon number conservation as a result of the SM field content and gauge, space-time symmetries. The core of this section is devoted to implanting this phenomena to the SISM in which the space-time symmetries are extended by SI. A quite predictive framework will be established, and in it the particle nature of DM such as mass origin, spin and interactions can be almost pined down. In what follows we will first demonstrate the idea using a toy model and then go to the realistic model.

\subsection{SISM with singlets: a toy model}\label{toy}

Let us start from a toy SISM that introduces only a few singlet scalars $S_i$ ($i=1,2,...n$) to implement the CW mechanism. Asides from the local symmetries $G_{\rm SM}=SU(3)_C\times SU(2)_L\times U(1)_Y$, the model respects Poincare and SI space-time symmetries. They restrict the most general renormalizable potential to be
\begin{align}\label{singlet}
-{\cal L}=\f{\ld_\phi}{2}(\Phi^\dagger\Phi)^2+\f{\ld_{ij}}{2}\Phi^\dagger\Phi
S_iS_j+\f{\ld_{ijkl}}{4!}S_iS_jS_kS_l,
\end{align}
The singlets obtain masses only through the second term, and thus it is convenient to work in the basis where they are diagonal, i.e, $\ld_{ij}=\ld_i\delta_{ij}>0$. Remarkably, an accidental $Z_2-$symmetry, only $S_i$ odd under it, emerges. The point is that SI forbids the cubic terms $S_iS_jS_k$, otherwise they would violate the accidental $Z_2$. If only the Higgs doublet acquires vacuum expectation value (VEV), $Z_2$ will survive after EWSB. Consequently the lightest singlet, denoted by $S$, will be stable and service as a DM candidate. Therefore, in the SISM which extends SM by singlets only, an accidental DM, a real singlet, can be accommodated. We will later argue that real singlet is the unique candidate and thus the following discussion on phenomenologies of $S$ actually yields generic features of aDM in the SISM.

On top of stability, other DM particle properties are largely specified. More exactly, the single term $\ld S^2|\Phi|^2/2$ accounts for them. Firstly, just like other massive members in the SM, DM obtains mass  $m_{\rm DM}=\sqrt{\ld/2}v$ after EWSB. Secondly, interactions between DM and the visible particles are via the Higgs portal. The Higgs mediated DM-nucleon spin-independent scattering has cross section $\sigma_{\rm SI}=4f_p^2{\mu_p^2}/{\pi}$, with $\mu_p$ the reduced mass of the proton-DM system and
\begin{align}\label{}
f_p&=
\f{\sqrt{\ld}}{2v}\f{m_p}{m_h^2}\L\sum_{q=u,d,s}
{f_{T_q}^{(p)}}+3\times\f{2}{27}{f_{T_G}^{(p)}}\R\approx4.5\times10^{-8} \ld^{1/2}\rm\,GeV^{-2},
\end{align}
where $m_h=125$ GeV. Values of the coefficients $f_{T_u}^{(p)}$, etc., can be found in Ref.~\cite{form,Gao:2011ka} (The updated data favors a smaller $f_{T_s}^{(p)}$, but it does not affect our ensuing qualitative conclusions):
\begin{align}\label{}
{f_{T_u}^{(p)}}=0.020 \pm 0.004,\quad {f_{T_d}^{(p)}}=0.026\pm 0.005, \quad  {f_{T_s}^{(p)}}=0.118\pm 0.062,
\end{align}
with ${f_{T_G}^{(p)}}=1-\sum_{q=u,d,s} {f_{T_q}^{(p)}}$. Note that $f_p$ depends on only one unknown parameter $\ld$, which allows one to derive a conservative upper bound $\ld\lesssim 0.03$ using data from one of the most stringent direct detection experiments, XENON100~\cite{XENON100}. It means that $m_{\rm DM}\lesssim 30$ GeV and then the Higgs invisible decay into a pair of $S$ kinematically opens and has a width
\begin{align}\label{inv}
\Gamma(h\ra SS)=\f{1}{32\pi}\f{\ld^2 v^2}{m_h}\L1-2\ld v^2/m_h^2\R^{1/2}.
\end{align}
The width of SM-like Higgs boson at 125 GeV is about 4.1 MeV. Even if the branching ratio of invisible decay is allowed to be as large as $20\%$~\footnote{This bound is based on the paper~\cite{Belanger:2013kya}. It is valid only in the case that  Higgs exotic decay opens while other couplings of Higgs boson are not modified, which is just the case in our paper.}, it still yields a more stringent upper bound on $\ld$ (than the one from XENON100) in turn DM mass:
\begin{align}\label{upper}
\ld\lesssim 0.013 \Rightarrow m_{\rm DM}\lesssim 20.0\rm\,GeV.
\end{align}
Therefore, the third feature of $S$ is that it should be in the relatively light region. Bear in mind that in the toy model $\ld$ controls all the relevant interactions of $S$, that small coupling causes $S$ to annihilate ineffectively, thus failing to have correct relic density $\Omega_{\rm DM}h^2\simeq0.1$. We should go beyond the toy model.

\subsection{Pinning down aDM model}

First of all, we argue that singlet scalar is the unique candidate of aDM. For instance, a fermionic singlet $\psi$ fails because SI can not forbid the coupling $\bar\ell \Phi\psi$ that negates an accidental $Z_2$. For non-singlets, one can arrive the same conclusion by taking into account the current experimental results. DM candidate with full electroweak charge dwelling in a representation of $SU(2)_L\times U(1)_Y$, the multiplet $(2j+1,Q_Y)$. Here $j$ is an integer or half integer, and $Q_Y$, for a given $j$, is chosen so that there is a neutral component in this multiplet. Concretely, its $2j+1$ components respectively have charges
\begin{align}\label{inv}
-j+Q_Y,~-j+1+Q_Y,~...,~j+Q_Y.
\end{align}
For instance, the SM Higgs doublet has $j=1/2$ and $Q_Y=-1/2$. For the lower $1/2\leq j\leq1 $, there are six potential multiplets accommodating a neutral component, $(2,1/2)_{f/s}$, $(3,0)_{f/s}$ and $(3,1)_{f/s}$ with $f/s$ denoting a fermion/scalar~\footnote{We do not distinguish the two multiplet having opposite signs of $Q_Y$ since they can be related by a conjugate.}. All of them can not accommodate an accidental $Z_2$ symmetry. $(2,1/2)_{f/s}$ share the same quantum numbers of the Higgs and lepton doublet $\ell$, so they can mix to spoil $Z_2$. The rest allows the following $Z_2-$violating operators 
\begin{align}\label{inv}
\bar \ell (3,0)_s\ell,\quad \bar \ell (3,0)_f\Phi,\quad \bar \ell^c (3,1)_s\ell,\quad  \bar \ell^c {(3,1)_f}\Phi .
\end{align}
Although the multiplets with even higher $j$ accommodate an accidental $Z_2$, they are ruled out by some phenomenology considerations, e.g., the $Z-$boson mediated DM-nucleon scattering has a too large rate. But this argument has a loophole, i.e., the multiplets with $Q_Y=0$ have no DM-DM-$Z$ coupling.

A more robust argument for the no-go comes from checking mass origins of the multiplets in the SISM, where the Higgs doublet VEV is the only, or at least the dominant, source of mass. For the neutral component of a scalar multiplet being DM, the arguments in the toy model directly apply: Higgs mediated DM-nucleon recoil forces coupling between the multiplet and Higgs doublet with a strength $\lesssim {\cal O}(0.01)$, which means that DM along with other members in the multiplet should have masses below about 30 GeV. As a consequence, the invisible decay width of $Z-$boson into the charged partners of DM turns out to be too large~\footnote{One may oppose that by adding operators $(H^\dagger T_H^a H)(\Phi^\dagger \tau^a \Phi)$, which however produce mass splittings $\propto Q^3_H v^2/m_{\rm DM}$, with $Q_H^3$ the charge of $H-$components under $T_H^3$. Thus some components will become even lighter.}. For the fermionic DM similar arguments are available. It should be a Dirac particle (denoted as $N^0$) because it must carry hypercharge, otherwise it can not couple to the Higgs doublet. Then, its mass comes from a term like $\ld_N \bar N^0N^0\Phi^0$. This means that Higgs again mediates DM-nucleon recoil and thus similarly the mother multiplet must contain intolerably light charged components. In conclusion, real scalar singlet is the unique viable aDM candidate. 
  
It is not the end of the story, and we can further pick out the EWSB trigger candidate. In the toy model aDM has a too small annihilation rate, as means that we have to cure this problem by introducing proper EWSB triggers (bear in mind that they are scalars) that open new effective annihilation channels for aDM. In principle, annihilations can proceed either at tree level into the SM fermions or at loop level into a pair of gluons/photons via a charged loop. But the latter requires a quite large quartic coupling between DM and triggers, thus jeopardizing perturbativity of the model at the weak scale. What is more, the indirect DM detections like gamma-ray line or antiproton search have already excluded such kind of light DM. Therefore, the trigger has to open tree level annihilation modes, into the light SM fermions. Alternatively, the trigger participates in spontaneously breaking of SI and provides some new light states, such as into which DM can annihilate. Obviously, extra Higgs doublets are good candidates because they can couple to fermions and participate in EWSB, potentially realizing both of these possibilities for DM annihilating. Further, one can argue that actually they are the unique candidates.

The argument is based on Higgs data. The QCD/QED charged triggers have large quartic couplings to $\Phi$ and at the same time they gain masses from those couplings, so they are likely to affect the Higgs production/decay rates too significantly. For example, the $(2j+1,Q_Y)$ trigger shifts the amplitude of Higgs to di-photon by an amount (normalized to the $W-$loop amplitude)~\cite{Guo:2013iij}, 
\begin{equation}\label{shift}
\delta_W =\f{1}{24}\times\f{7}{8}\sum_{n=1,...,2j}(j-n+Q_Y)^2. 
\end{equation}
Only the multiplet with $j\leq1$ can change the Higgs to di-photon rate by less than $40\%$. In particular, for $(2,\pm1/2)$ the corresponding change is about $7\%$. While the widely used triplets $(3,0)_s$ and $(3,\pm1)_s$ recurs $Z_2-$violating respectively via $\Phi^\dagger (3,0)_s\Phi S$ and $\Phi (3,1)_s\Phi S$, along with the first and third operators listed in Eq.~(\ref{inv}). Note that the multiplets with peculiar quantum numbers (e.g., $Q_Y=1/10$) can not couple to the SM particles in the form of $Z_2-$breaking, and thus they are stable charged relics which are definitely excluded.

Now further using the economical criteria, we eventually are able to pine down the model, two-Higgs-doublets plus a real singlet (2HDM+$S$). There are different versions of 2HDM classified by the pattern of couplings between the extra doublet $\Phi'$ (which has identical SM quantum numbers with $\Phi$) and SM fermions~\cite{Branco:2011iw}. For definiteness, here we focus on type-II, where $\Phi$ and $\Phi'$ (in order to follow the commonly used convention hereafter we define $\Phi_2\equiv \Phi$ and $\Phi_1=\Phi'$) couple to the up-type and down-type quarks (and leptons), respectively: 
\begin{align}\label{yukawa:II}
-{\cal L}_{\rm II}= y^u\overline{ Q_L}\wt \Phi_2 u_R+y^d\overline{ Q_L}\Phi_1 d_R+ y^e \overline{ l_L} \Phi_1e_R+h.c.
\end{align}
As usual, $\wt\Phi_2 \equiv i \sigma_2 \Phi_2^*$.  Both doublets $\Phi_{1,2}$ are supposed to develop VEVs to account for the fermion masses. We define $\tan \beta\equiv\langle \Phi_2\rangle/ \langle
\Phi_1\rangle=v_2/v_1$. The most general tree-level Higgs potential, restricted by SI, takes the form of
\begin{align}
 V_{\rm{2HDM}} &= 
\frac{1}{2}\lambda_1\left(\Phi_1^\dagger\Phi_1\right)^2+\frac{1}{2}\lambda_2\left(\Phi_2^\dagger\Phi_2\right)^2 
 +\lambda_3\left(\Phi_1^\dagger\Phi_1\right)\left(\Phi_2^\dagger\Phi_2\right)+\lambda_4\left(\Phi_1^\dagger\Phi_2\right)\left(\Phi_2^\dagger\Phi_1\right)
\cr
&+\Big\{ \frac{1}{2}\lambda_5\left(\Phi_1^\dagger\Phi_2\right)^2+\Big[\lambda_6\left(\Phi_1^\dagger\Phi_1\right) 
+\lambda_7\left(\Phi_2^\dagger\Phi_2\right)\Big]\left(\Phi_1^\dagger\Phi_2\right) + \rm{h.c.}\Big\}. 
\label{eq:2hdmgen}
\end{align}  
Even if all $\ld_{i}$ are assumed to be real, it still contains seven parameters. So we want to make some reasonable simplifications. Note that the $\ld_{1-4}-$terms conserve two global Abelian symmetries $U(1)_1$ and $U(2)_2$, under which only $\Phi_1$ and $\Phi_2$ are charged, respectively. As a result, if $\ld_{5-7}$ all are vanishingly small, an pseudo Goldstone boson will emerge. This light particle will be very helpful in DM annihilating. We as usual turn off $\ld_{6,7}$. The left five parameters are minimally required in radiative EWSB, and can not be further reduced. The final part of the model is about the dark matter field $S$, 
\begin{align}\label{doublet}
-{\cal L}_S=\sum_{i=1,2}\f{\eta_{i}}{2}S^2|\Phi_i|^2+{\eta_{12}}S^2{\rm Re}( \Phi_1^\dagger
\Phi_2) +\f{\eta}{4!} S^4,
\end{align}
where $\eta$ actually is irrelevant. The above three equations constitute the complete Lagrangian of the scale invariant 2HDM$+S$. In the following subsection we will investigate radiative EWSB in detail.

\subsection{Radiative EWSB via Gilard-Weinberg (GW) approach}

Here EWSB through the CW mechanism involves two VEVs $v_1$ and $v_2$, and thus the situation is a little bit different to the general analysis made in Section~\ref{CW:ge} where only one VEV is involved. The Gilard-Weinberg (GW) approach~\cite{GW} should be adopted to handle this situation. For later use, we decompose the fields in component as 
\begin{align}\label{}
\Phi_i^T=\L(R_i+iI_i)/\sqrt{2},\,H_i^-\R.
\end{align}
In the physical vacuum, $R_i$ are supposed to acquire VEVs. It is illustrative to rewrite Eq.~(\ref{eq:2hdmgen}) in terms of the above decomposition. Then, the tree level Higgs potential relevant to vacuum determination is given by
\begin{align}\label{}
V(R_1,R_2)=\f{\ld_1}{8}R_1^4+\f{\ld_2}{8}R_2^4+\f{\ld_3+\ld_4+\ld_5}{4}R_1^2R_2^2.
\end{align}
The other part involves other components, who gain masses via terms like $R_i^2 |H_i^-|^2$. In order to determine the vacuum at quantum level, one should use the classical background fields $\phi_{{\rm cl},i}$ instead of $R_i$ as field variables. Thus as usual we make the shifts $R_i\ra \phi_{{\rm cl},i}+R_i$. Then, all fields, including the SM fields, gain $\phi_{{\rm cl},i}-$dependent mass terms. Later, they will be used to evaluate the CW potential. 
 
Now we follow the GW procedure to investigate EWSB in the 2HDM with SI. First of all, one should figure out the tree level vacuum from the tadpole equations $\partial V(\phi_{{\rm cl},1},\phi_{{\rm cl},2})/\partial \phi_{{\rm cl},i}=0$. But,  as a result of SI, the solutions lead to an extremum line instead of point of the Higgs potential. Such an extremum line is dubbed as a flat direction in the two-dimensional field space, i.e., $\vec \phi_{\rm cl}=(\phi_{{\rm cl},1},\phi_{{\rm cl},2})\equiv \phi_{\rm cl}\,\vec n$ with $\vec n=(n_1,n_2)$ satisfying  
\begin{align}\label{tadpole}
\f{n_2^2}{n_1^2}=\f{\phi_{{\rm cl},2}^2}{\phi_{{\rm cl},1}^2}=-\f{\ld_1}{\ld_3+\ld_4+\ld_5}=-\f{\ld_3+\ld_4+\ld_5}{\ld_2}.
\end{align}
As one can see, $\vec n$ can be determined unambiguous while $\phi_{\rm cl}$ is not determined (otherwise spontaneously breaking of SI is achieved at tree level). We solve the tadpole equations by expressing $\ld_{1,2}$ in terms of others, and further eliminate them from the potential. Then, it is straightforward to derive the background fields dependent mass squared of the Higgs bosons:
\begin{align}\label{mh2}
m_{H^-}^2=&-\f{\ld_{4}+\ld_5}{2} {\phi_{\rm cl}^2}, 
\cr
 m_{H}^2=&-\L\ld_{3}+\ld_4+\ld_5\R{\phi_{\rm cl}^2},\cr
 m_{A}^2=&-\ld_5 {\phi_{\rm cl}^2}.
\end{align}
We have not written out the conventional Goldstone bosons which are eaten by the electroweak gauge bosons. In particular, as before  there is a massless CP-even Higgs boson $h$ as a result of the SI spontaneously breaking. Additionally, the SM massive gauge bosons and as well top quarks also have $\phi_{\rm cl}-$dependent masses
\begin{align}\label{SMC}
m_{W}^2=&\f{g_2^2}{4}\phi_{\rm cl}^2, \quad 
 m_{Z}^2=\f{g_2^2}{4\cos^2\theta_w}\phi_{\rm cl}^2,\quad
 m_{t}^2=y_t^2 {\phi_{\rm cl,1}^2}.
\end{align}
Unlike others, the top quark masses depend on $\phi_{\rm cl,1}$ instead of $\phi_{\rm cl}$.

We pause to make several important comments. Firstly, the mixing angles of all the three Higgs mass squared matrices are determined by the flat direction, up to a sign. Concretely speaking, for the charged and CP-odd Higgs bosons we have
\begin{align}\label{}
G^-=&\cos\beta  H_1^-+\sin\beta H_2^-,\quad   H^-=-\sin\beta H_1^-+\cos\beta  H_2^-,\cr
G^0=&\cos\beta  I_1+\sin\beta \,I_2,\quad ~~~~~~~  A=-\sin\beta I_1+\cos\beta \, I_2,
\end{align}
with $\tan\beta\equiv \phi_{\rm cl,2}/\phi_{\rm cl,1}=\pm n_2/n_1$. As for the CP-even Higgs bosons we have $h=-\sin\alpha \,R_1+\cos\alpha R_2$ and $H=\cos\alpha  R_1+\sin\alpha R_2$, with
\begin{align}\label{mixing}
\sin\alpha=\f{n_1}{\sqrt{n_1^2+n_2^2}}=\cos\beta,
\end{align}
Thus we have $\alpha=\pi/2+\beta$, which is a crucial difference to the usual 2HDM where $\alpha$ somehow deviates from that value. As a consequence of Eq.~(\ref{mixing}), $h$ has full SM coupling to vector bosons while $H$ exactly decouples from them. So, the PGSB $h$ will be the SM-like Higgs boson. Secondly, as expected, the CP-odd Higgs boson mass squared $ m_{A}^2\propto \ld_5$, which indicates the explicitly breaking of $U(1)_1$ and $U(1)_2$ (to the global correspondence of $U(1)_Y$). This feature will be crucial in dark matter phenomenology. On the other hand, in order to suppress the Higgs exotic decay $h\ra AA$, we should impose the lower bound on $A$ mass, $m_A>m_h/2$ (thus $\ld_5\gtrsim 0.1$). The vertex $h-A-A$ has a large coupling $\mu_{hAA}=-(\ld_3+\ld_4-\ld_5)v\approx \ld_3 v$, which, with value of $\ld_3$ that will be determined in Eq.~(\ref{ld:para}), is around 600 GeV. Consequently, once $h\ra AA$ kinematically opens, it will overwhelmingly dominate Higgs decay. Therefore, that light $A$ is definitely ruled out by Higgs data. Last but not the least,  the Higgs spectrum is well split for generic $\ld_{3,4,5}$, thus violating the custodial $SU(2)$ symmetry. The electroweak oblique corrections to the $S,T$ and $U$ parameters~\cite{STU} yield a strong exclusion, except for accidental degeneracy between $H^\pm$ and $H$, or $A$. Viewing from dark matter phenomenology, we are interested in the former case $m_{H^\pm}\approx m_H$, which implies a relationship between the quartic couplings:
\begin{align}\label{degen}
\ld_4\approx -2\ld_3.
\end{align}
It is obtained for a negligible $\ld_5\sim 0.1$ compared to $\ld_{3,4}$, which will be justified later.

We resume the discussion on EWSB at loop level, which determines the value of $\phi_{\rm cl}$ and moreover modify the CP-even Higgs sector. With Eq.~(\ref{mh2}) and Eq.~(\ref{SMC}) at hand, we can calculate the radiative correction to the tree level potential, i.e., the CW potential $V_{\rm CW}$ along the flat direction. It again takes the form of Eq.~(\ref{veff}) with
\begin{align}\label{}
B\approx\f{1}{64\pi^2}\L2\times\f{(\ld_4+\ld_5)^2}{4}+ {\L\ld_3+\ld_4+\ld_5\R^2}+ \ld_5^2+6(g_2/2)^4+3(g_2/2\cos\theta_w)^4-12\times \f{\sin^4\beta y_t^4}{4}\R.
\end{align}
$A$ is similarly obtained. For sufficiently large $\ld_{3}$ and/or $\ld_{4,5}$, $B$ is positive. From the new extreme condition for the total potential $V(\phi_{{\rm cl},1},\phi_{{\rm cl},2})+V_{\rm CW}$ one can find that the minimum  locates at
\begin{align}\label{}
\langle\phi_{\rm cl}\rangle \equiv v=Q \exp\L- A/2B-1/4\R\approx 246\rm\,GeV. 
\end{align}
It practically establishes a relation among the quartic couplings once $Q$ is chosen, a consequence of the dimensional transmutation. The PGSB $h$ gains mass squared $8B\phi^2_{\rm cl}$, which according to Eq.~(\ref{SIPGSB}) should be near $(\rm125 GeV)^2$. Therefore, with the help of Eq.~(\ref{degen}) we can fix the quartic couplings 
\begin{align}\label{ld:para}
\ld_3\approx 2.40,\quad \ld_4\approx -4.80,
\end{align}
while $\ld_1\ll1$. As a prediction, the model presents two heavier Higgs bosons $H^{\pm}$ and $H$ having almost regenerate masses at 381 GeV. Moreover, from the tree level tadpole conditions Eq.~(\ref{tadpole}), $\ld_{1,2}$ are determined to be
\begin{align}\label{}
\ld_1= \ld_3 \tan^2\beta\approx 2.40*\tan^2\beta,\quad \ld_2\approx 2.40/\tan^2\beta.
\end{align}
A large $\tan\beta$ is thus disfavored because it blows up $\ld_1$ and hence violates tree level unitarity. All of the above results are obtained in the approximation by ignoring loop corrections to the tree level mass spectrum, except for the PGSB. A more complete analysis is already employed in Ref.~\cite{Lee:2012jn}. It includes all radiative corrections which could lead to a sizable correction to the mixing angle of CP-even Higgs bosons (but not of the charged and CP-odd Higgs bosons). And their study shows that, if $\tan\beta\lesssim2$ and $m_{H^+}\approx m_H\sim 400$ GeV, a 125 GeV SM-like Higgs boson can be accommodated, satisfying all experimental and theoretical constraints. This is consistent with our approximate analysis.

As a mention, in our scenario under consideration, all the new Higgs states are allowed by the current collider data. The most sensitive probe to them is from the CMS experiment searching for the heavy neutral Higgs boson in the MSSM decaying into a pair of $\tau$. It collects data corresponding to an integrated luminosity of 24.6 fb$^{-1}$~\cite{Khachatryan:2014wca}. The results, demonstrated in the $m_h^{\rm max}$ scenario, can be applied to 2HDM in the decoupling limit. $H/A$ around 400 GeV has not been touched yet for $\tan\beta\sim2$. As for the light CP-odd Higgs boson $A$, in the region 63-100 GeV that will be considered later, colliders also fail in hunting for this fully electroweak particle. At LEP, it is dominantly produced via the processes $e^+e^-\ra  Z^*(W^*)\ra H(H^\pm) A$. However, the mass of $H(H^\pm)$ exceeds the threshold of LEP and thus the process is not kinetically accessible. At LHC that light particle is buried in the huge QCD backgrounds. As a matter of fact, bounds on the new Higgs bosons with mass below 100 GeV are even not presented in Ref.~\cite{Khachatryan:2014wca}. However, flavor physics may impose strong bounds. For instance, $b\ra s\gamma$, with updated NNLO QCD predictions, yields lower bound on $H^\pm$ mass in this model $m_{H^+}>480$ GeV at 95$\%$ C.L. and  $m_{H^+}>350$ GeV at 99$\%$ C.L.~\cite{Misiak:2015xwa}. 
Thus our prediction $m_{H^+}\approx 381$ GeV has been excluded at 95$\%$ C.L. Taking it seriously, we may have to modify the Yukawa structure, e.g., the second Higgs doublet merely couples to leptons. Anyway, it does not affect most of the discussions of this paper, where the Higgs sector plays the main role.

\subsection{Saving the light accidental dark matter by a light $A$}

In the toy model with a single Higgs doublet $\Phi_2$, aDM with correct relic density is already excluded by the DM direct detection. Potentially, the presence of $\Phi_1$ is able to save it in two different ways. But restricted in the type-II 2HDM we need fine tuning more or less in both way.

\begin{figure}[htb]
\includegraphics[width=6.5in]{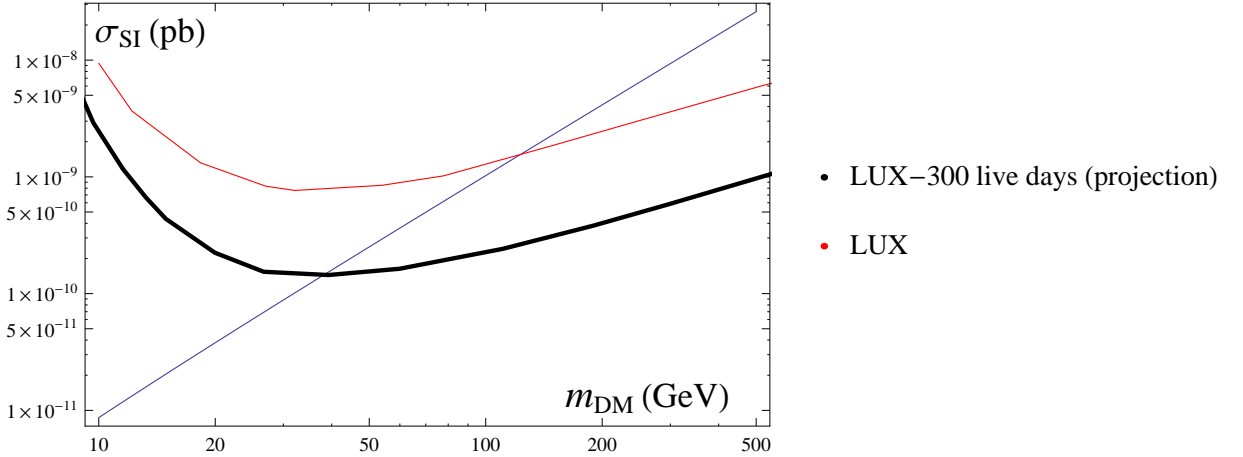}
\caption{Given a smaller $\tan\beta$, DM direct detection bound (LUX, red line) allows for a heavier dark matter near 100 GeV in the $\Phi_1-$portal. The projected LUX-300 live days (black line) of WIMP search is able to fully cover the surviving parameter space.}\label{DM:heavy}
\end{figure}

Before heading towards the details, we show that  in the $\tan\beta\sim1$ scenario a crucial difference between 2HDM$+S$ and the toy model described in Section~\ref{toy} arises. Concretely, in this scenario DM can become relatively heavy, much heavier than the previous bound given in Eq.~(\ref{upper}). The point is that $\Phi_1$ develops VEV around 100 GeV and moreover accommodates a CP-even Higgs boson $H$, which is much heavier than $h$; As a consequence, now the aDM $S$ can acquire a large mass via the $\eta_1-$term in Eq.~(\ref{doublet}) without rendering an intolerably large $\sigma_{\rm SI}$. Let us explicitly show this. Here aDM mass receives several contributions:
\begin{align}\label{}
m_{\rm DM}^2=\f{v^2}{2}\L{\eta_1}\sin^2{\beta}+ {\eta_2}\cos^2{
\beta}+\eta_{12}\sin2{\beta}\R.
\end{align}
The $\eta_2-$term is from the conventional term $\f{1}{2}\eta_2S^2|\Phi_2|^2$, and $\eta_2\ll1$ is required to suppress  $\sigma_{\rm SI}$. Hence, for the moment we just neglect this parameter and derive the Higgs-DM-DM couplings,
\begin{align}\label{HSS}
-{\cal L}_{in}\supset &\f{1}{2}\L\f{{\eta_1}}{2}\cos\alpha \cos\beta + \eta_{12}\L\cos\beta\sin\alpha+\sin\beta\cos\alpha\R\R v \,HS^2+\f{1}{2} \eta_{12}\L\cos\beta\cos\alpha-\sin\beta\sin\alpha\R v \,hS^2\cr
\equiv &\f{1}{2}\mu_H HS^2+\f{1}{2}\mu_h hS^2.
\end{align}
We have dropped the $\eta_1\sin^2\alpha-$contribution to the coupling of $hS^2$. As one can see, to decouple  $h$ from DM (or, sufficiently suppress their coupling), either $\eta_{12}\ll1$ or $\tan\beta\gg1$ is required. The latter is inconsistent with the 125 GeV SM-like Higgs boson and therefore we have to consider a sufficiently small $\eta_{12}$. To estimate the cross section of $H-$mediated DM-nucleon scattering, we further need its coupling to fermions,
\begin{align}\label{Hff}
-{\cal L}_{in}\supset \f{\cos\alpha}{\cos\beta}\f{m_d}{v} H\bar d d+\f{\cos\alpha}{\cos\beta}\f{m_e}{v} H\bar e e
+\f{\cos\alpha}{\sin\beta}\f{m_u}{v} H\bar u u,
\end{align}
with family indices buried. Here we do not use the tree level value of $\alpha=\pi/2+\beta$ and assume that radiative correction will make $\sin\alpha\ll1$ (thus $\cos\alpha\approx$1) even for a relatively small $\tan\beta\lesssim2$~\cite{Lee:2012jn}. Having collected all the relevant terms, now it is ready to write $\sigma_{\rm SI}$ as 
\begin{align}\label{}
\sigma_{\rm SI}\approx 0.08\times\f{\eta_1}{v^2} \L\f{\cos^4\alpha}{\sin^2\beta}\R\L\f{\sqrt{m_p\mu_p}}{m_H}\R^4.
\end{align}
To get it we have worked in the $\Phi_1-$portal limit by turning on $\eta_1$ only. Because of the heaviness of $m_H$, now $\eta_1$ can be order 1 number and $m_{\rm DM}$ is allowed to be as heavy as 100 GeV, see Fig.~\ref{DM:heavy}.

\begin{figure}[htb]
\includegraphics[width=5.0in]{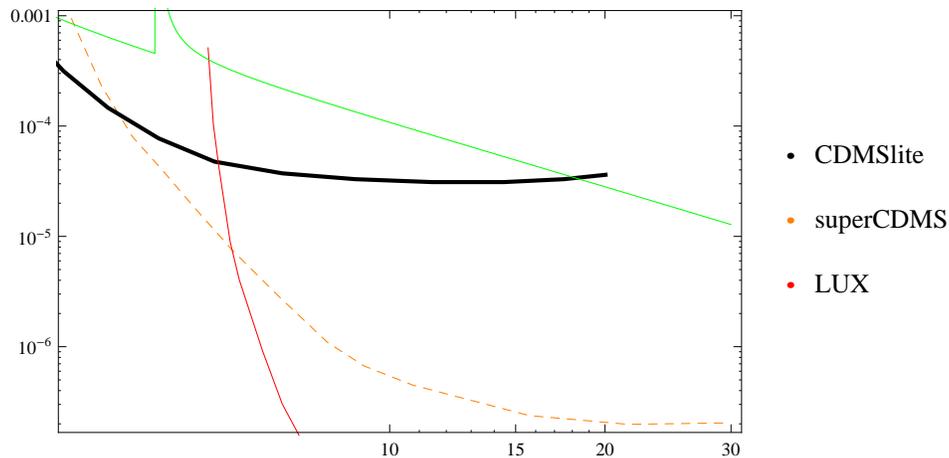}
\caption{Exclusion of the light aDM (green line) in the type-II 2HDM+$S$ on the $m_{S}(\rm GeV)-\sigma_{\rm SI}(\rm pb)$ plane. Constraints are from CDMSlite (black line), superCDMS (dashed orange line) and LUX (red line).}\label{DM}
\end{figure}

One can show that aDM in the low mass region (far below $m_h/2$) actually is already excluded, if the light DM can only annihilate into a pair of down-type quarks or leptons via $H-$mediation. The cross section is given by
\begin{align}\label{}
\sigma v_{f\bar
f}&=\L\f{\cos\alpha}{\cos\beta}\f{\mu}{v}\R^2N_f\f{1}{4\pi}\f{m_f^2}{m_{H}^4} \L
1-m_f^2/m_{\rm DM}^2\R^{3/2},
\end{align}
with $N_f$ the color factor, taking a value 3 for quarks and 1 for leptons. With the help of Eq.~(\ref{HSS}), the term in the first bracket can be simplified to $\cos^2\alpha\, \eta_1/2\ra  \eta_1/2$ if $\eta_{12}\ra0$; Oppositely, if $\eta_1\ra0$ it is simplified to $\eta_{12}\tan\beta$, which shows the $\tan\beta$ enhancement while the former case does not. Regardless of the undetermined values $\tan\beta$ and $\cos\alpha$, we can rule out the scenario of $\Phi_1-$portal into the light fermions. Consider the ratio
\begin{align}\label{ratio} \f{\sigma_{\rm
SI}}{\sigma v_{f\bar f}}\approx\f{0.04}{N_f}\times\L\f{m_p\mu_p}{m_{\rm
DM}m_f}\R^2 \L 1-\f{m_f^2}{m_{\rm DM}^2}\R^{-\f{3}{2}},
\end{align}
which, for a given fermion $f$, takes a fixed value up to DM mass. From it we see that: For $m_{\rm DM}\gtrsim m_b$, DM dominantly annihilates into a pair of $b$ and then typically $\sigma_{\rm SI}$ should be larger than $10^{-4}({ \rm GeV}/m_{\rm DM})^2\sigma v_{f\bar f}$. So, even DM is around 100 GeV, the resulting $\sigma_{\rm SI}$ is still at least $10^{-8}$ pb. Clearly, LUX excludes this region; For the even lighter DM, $m_\tau\lesssim m_{\rm DM}\lesssim m_b$, one gets $\sigma_{\rm SI}\sim10^{-3}$ pb. CDMSlite~\cite{Agnese:2013jaa} along with SuperCDMS~\cite{Agnese:2014aze}, which have fairly low threshold and thus are sensitive to low mass WIMP, rule it out. The situation is shown in Fig.~\ref{DM}. 

\begin{figure}[htb]
\includegraphics[width=3.3in]{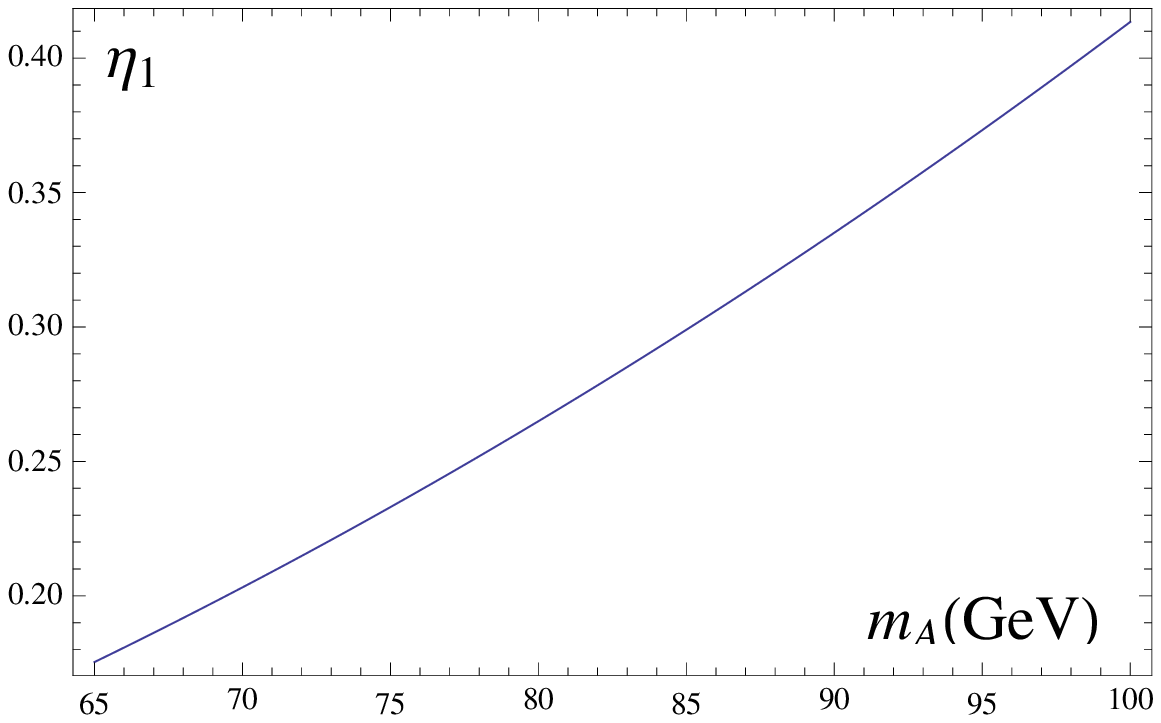}
\includegraphics[width=3.3in]{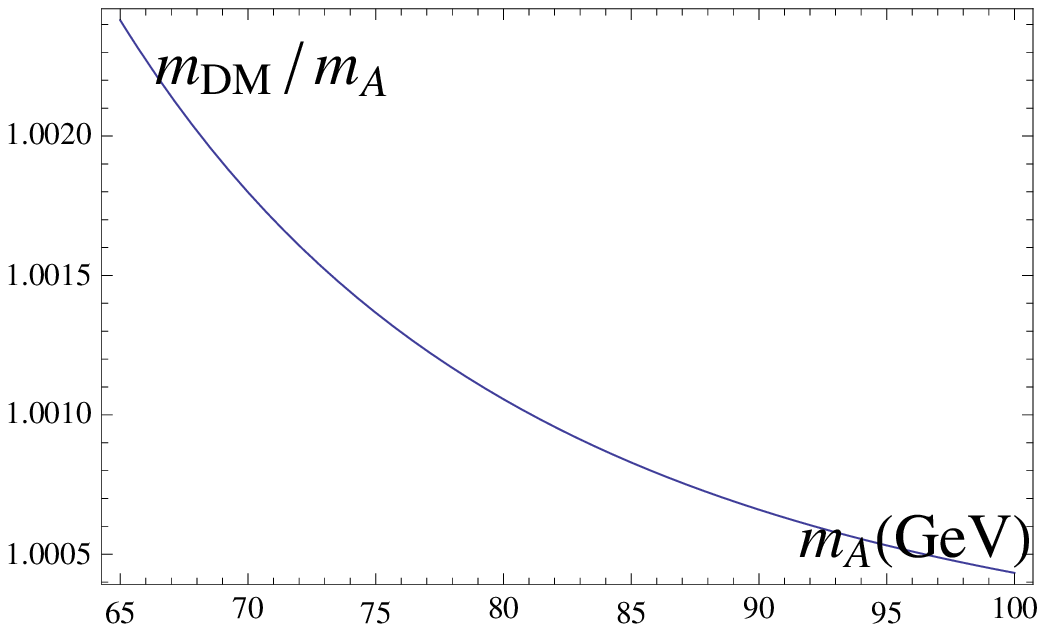}
\caption{Scenario of forbidden annihilating channel $SS\ra AA$. Left: Required $\eta_1$ for a given $m_A$; Right: The required mass degeneracy between $m_{\rm DM}$ and $m_A$.}\label{AA}
\end{figure}

Now it is at the position to demonstrate how a relatively heavy DM (close to or above $m_h/2$) can annihilate away, safely but at the price of tuning. The first way is trivial, falling back on the DM coupling to $h$, i.e., $hS^2$, that we have neglected before. Consider the annihilation $SS\ra b\bar b$ mediated by $h$ instead of $H$ in the $s-$channel. Now $m_{\rm DM}$ can be near the resonant pole of the Higgs boson $m_h/2\approx 63$ GeV, so this channel can always have a cross section $\simeq1$ pb, even for a quite small $\mu_h$ in order to avoid the LUX bound. On the other hand, $\sigma_{\rm SI}$ from $H-$mediation is a few $10^{-10}$ pb, which can be covered 
by the projected LUX-300 live days of WIMP search, see Fig.~\ref{DM:heavy}. The second way is more interesting. Consider the annihilation $SS\ra AA$, from the terms that generate mass for $S$, e.g., $S^2|\Phi_1|^2$. The relevant vertex can be easily derived, 
\begin{align}\label{}
-{\cal L}_{in}\supset\f{1}{4}\L\eta_1\sin^2\beta+\eta_2\cos^2\beta-\eta_{12}\sin2\beta\R S^2A^2.
\end{align}
Actually, even a single parameter $\eta_1$ is able to produce correct DM relic density. To see this, we turn off $\eta_{2}$ and $\eta_{12}$ and calculate the cross section,
\begin{align}\label{}
\langle\sigma v\rangle_{AA}\approx\f{1}{64\pi}\f{\eta_1^2}{m_S^2}\L 1-\f{m_A^2}{m_{\rm DM}^2}\R^{\f{1}{2}}=\f{1}{32\pi}\f{\eta_1}{v^2\sin^2\beta}\L 1-\f{m_A^2}{m_{\rm DM}^2}\R^{\f{1}{2}}.
\end{align}
As once can see, if the phase space factor is negligible (with $\tan\beta=2$), then one gets $\eta_1\approx 0.012$. As a result, DM has mass 17 GeV only, much below $m_h/2$. Recalling that $m_A$ is required to lie above $m_h/2$  (as a reminder, to forbid the large decay $h\ra AA$), thus that annihilation actually is forbidden. To shift DM to the region above $m_h/2$, the phase space factor is supposed to be highly relevant. We can understand it as this: The $SS\ra AA$ channel, which has a cross section proportional to $\eta_1$, is too effective for the heavier DM; In order to allow a larger $\eta_1$ (thus heavier $m_{\rm DM}$), we must count on the substantial suppression from the phase space  factor. To see this more clearly, in Fig.~\ref{AA} we plot the solution that leads to correct relic density on the $m_A-\eta_1$ plane and $m_A-m_{\rm DM}/m_A$ plane. From them it is seen that $\eta_1$ should be ${\cal O}(0.1)$, and the degeneracy between DM and $m_A$ is fairly high, at order ${\cal O}(0.001)$~\footnote{Our estimation is based on the cross section with exact non-relativistic limit of DM, without considering thermal effect. This effect will modify the prediction qualitatively~\cite{forbid},  but the required degeneracy will not be changed much.}. The resulting phase space suppression is about a few percents~\footnote{The three-body annihilation mode $SS\ra Ab\bar b$, which opens for $m_{\rm DM}\approx m_A$ and is not suppressed by phase space, may matter. But numerically it is not important, because this mode is additionally suppressed by small Yukawa couplings and $1/(2\pi)^3$. As a conservative estimation, the suppression is $\sim (m_b\tan\beta/v)^2 /(2\pi)^3\sim 10^{-5}$, much more sever than the suppression considered here. }. We would like to stress that this scenario, due to the $H-$mediated DM-nucleon recoil, is also covered by the next round of LUX. In summary, our aDM, within the 65-100 GeV region, can be seen or ruled out in the near future.

To end up this section, it is of interest to comment aDM in other types of 2HDM such as the lepton-specific type. In this type $\Phi_1$ only couples to leptons and thus it becomes a leptonic portal, given a suppressed mixing between the two CP-even Higgs bosons. As a consequence, the ratio Eq.~(\ref{ratio}) does not hold and then we can get a viable aDM easily. Besides, it is of interest to note that a light DM with large $\sigma_{\rm SI}$, predicted in our model, is required to explain the CoGeNT anomaly, which hinted a 8 GeV DM with $\sigma_{\rm SI}\sim 10^{-5}$ pb~\cite{CoGeNT}. Despite of the inconsistence with LUX (and others) in the conventional models, 2HDM$+S$ with a spectator $\Phi_1$ may reconcile them. The point is that $\Phi_1$ couples to quarks in a well temped way such that isospin-violating DM can be accommodated~\cite{Gao:2011ka}.

\section{Conclusions}

Extending space-time symmetries by SI may provide a simple way to address the Higgs naturalness problem. With such this extended symmetries, we attempt to embed accidental DM into the SISM. It is found that 2HDM$+S$ is the unique model that can give rise to acceptable aDM phenomenologies. We study the the case with a type-II 2HDM, which presents two predictions: A real scalar DM near or above the Higgs pole and heavy Higgs states about 380 GeV. They can be examined soon both from DM detections and LHC searches. Because of its simplicity, the real singlet scalar DM has been extensively studied based on the SM~\cite{realsinglet} or 2HDM~\cite{Cai:2013zga}. Here we reveal that this somewhat trivial particle actually has depth: It is the unique aDM candidate by SI and therefore is indirectly related to the solution to the Higgs naturalness problem.

In this work we consider the 2HDM potential with $\ld_{6,7}=0$, which yields a strong tadpole condition Eq.~(\ref{tadpole}), i.e., $\tan\beta$ can not be very large otherwise $\ld_1$ blows up. Thus, it is of interesting to investigate the case with at least one non-vanishing $\ld_{6,7}$, where the bound on $\tan\beta$ may be relaxed. In addition, beyond the WIMP scenario such as in the feebly interacting massive particle as DM scenario, the model building for aDM will become sharply different~\cite{Kang:2015aqa}.


\section{Acknowledgements}
We would like to thank P. Ko for helpful discussions. This research was supported in part by the China Postdoctoral
Science Foundation (No. 2012M521136) (ZK).

\end{document}